\documentclass{ptapap}
\usepackage{amsmath}
\usepackage{amssymb}
\usepackage{color}
\usepackage{xcolor}
\usepackage{hyperref}
\usepackage{subcaption}
\usepackage{graphicx}

\newenvironment{myfont}{\fontfamily{lmtt}\selectfont}{\par}
\DeclareTextFontCommand{\textmyfont}{\myfont}

\def\kms{\,km\,s$^{-1}$}

\def\hb{{\sc{H}}$\beta$\/}

\def\feii{Fe{\sc{ii}}}
\def\rfe{R$_{\rm{FeII}}$}

\def\mbh{$M\mathrm{_{BH}}$}

\def\mdot{$\dot{\mathcal{M}}$}

\def\LLEdd{$L\mathrm{_{bol}}/L\mathrm{_{Edd}}$}

\def\RL{$R\mathrm{_{H\beta}}-L_{5100}$}

\def\zsun{Z$_{\odot}$}
\def\un{$\log \rm{U} - \log \rm{n_{H}}$}

\author{Swayamtrupta Panda}[CFT, CAMK]

\affil[CAMK]{Nicolaus Copernicus Astronomical Center, Polish Academy of Sciences, Bartycka 18, 00--716 Warsaw, Poland}
\affil[CFT]{Center for Theoretical Physics, Polish Academy of Sciences, Al. Lotnik\'ow 32/46, 02--668 Warsaw, Poland}

\title{Explaining the Broad-line Region through photoionisation modelling}

\begin{document}

\maketitle

\begin{abstract}

Broad-line regions (BLR) are one of the main components that constitute the phenomenological picture of active galaxies near the vicinity of the accreting supermassive black holes. Both theoretical and observational studies have shown that the BLR is made of dense, ionized gas clumps that have a strong virialized distribution at parsec-scale distances from the nuclei. Using a theoretically motivated photoionized gas model, I constrain the ionisation parameter (\textit{U}) and cloud density (\textit{n$_{\rm{H}}$}) as a function of the strength of the \feii{} emission. Recent observations in the reverberation mapping studies have contested the standard radius-luminosity relation showing increased dispersion in the relation, in particular, after the inclusion of highly accreting quasars. I incorporate the departure coefficient that accounts for this dispersion. This departure term in terms of the dimensionless accretion rate (\mdot{}) and Eddington ratio (\LLEdd{}), also includes the virial factor that accounts for the BLR geometry. I then combine the fundamental plane relation for the BLR to connect the \feii{} strength (\rfe{}) in terms of \textit{U} and \textit{n$_{\rm{H}}$} using selected values for the shape of the broad \hb{} profile.

\end{abstract}

\section{Introduction}
\label{sub:intro}

The Broad-line region (BLR) in the active galaxies (AGN) has been extensively studied from the X-rays to the NIR regime during the last three decades \citep[see the reviews of][]{2000ARA&A..38..521S,2009NewAR..53..140G}. One of the most puzzling aspects of the line spectra emitted by the BLR is the \feii{} emission, whose numerous multiplets form a pseudo-continuum which extends from the UV to the optical region due to the blending of approximately 10$^5$ lines. This emission constitutes one of the most important contributors to the cooling of the BLR. 

I propose a novel method to combine our existing knowledge about the BLR from both the theory and the observational standpoints. This analysis will allow us to constrain the \un{} space that is generally used to describe the line emission using photoioization as the fundamental radiation mechanism, in terms of the strength of the \feii{}, i.e. \rfe{}\footnote{the integrated flux of the \feii{} between 4434-4684 \AA~ normalised by the \hb{} flux. See \citet{panda18b} (and references therein) for more details.}. I will then be able to connect these quantities to (i) the underlying accretion disk (i.e., in terms of the global accretion rate); (ii) a well-represented ionizing continuum for a source accreting about the Eddington limit ($L_{\rm{Edd}}$); and (iii) combining our knowledge of the BLR cloud composition and geometry. This analytical expression in its final form will be able to describe the BLR with more certainty, wherein observed sources with \rfe{} measurements can be projected to retrieve the information about the \textit{U} and \textit{n$_{\rm{H}}$}, and vice versa.

The paper is organised as follows. Section \ref{sub:methods} derives the \un{} relation in terms of the \RL{} relation \citep{bentz13} and subsequently in terms of Eddington ratio ($\lambda_{Edd}$ = \LLEdd{}) and black hole mass (\mbh{}). The subsequent steps are highlighted briefly in Sec. \ref{exten} which will be addressed in detail in a subsequent paper. The CLOUDY \citep{f17} setup for this model is described in Sec. \ref{sub:cloudy}. Section \ref{sub:results} describes and analyzes one basic scenario of comparing the constant density with the constant pressure model in terms of the \feii{}, \hb{} line emissions and correspondingly on \rfe{} in the \un{} space. I summarize with plans for immediate future work leading into a more comprehensive paper in Sec. \ref{sub:future}. 
\section{Methods and Analysis}
\label{sub:methods}

\subsection{Analytical description}
In order to realise the parameter space for the BLR and to link the physical quantities (U,$\rm{n_H}$) and the observable (i.e, \rfe{}), I derive an analytical relation as described in the following sub-sections.

\subsubsection{Derivation under \textit{R-L$_{5100}$} constraint}
Starting with the conventional description of the ionization parameter,\\
\begin{equation}
    U = \frac{Q(H)}{4\pi R_{BLR}^2 n_H c} = \frac{\Phi(H)}{n_{H} c}
\end{equation}
where, \textit{Q(H)} is the number of hydrogen-ionizing photons emitted by the central object (in $\mathrm{s^{-1}}$); \textit{R$_{BLR}$} is the separation between the central source of ionizing radiation and the inner face of the cloud (in cm); \textit{$\rm{n_H}$} is the total hydrogen density (in $\mathrm{cm^{-3}}$); and, \textit{$\Phi(H)$} is the surface flux of ionizing photons (in $\mathrm{cm^{-2}\;s^{-1}}$ ).\\

The \textit{Q(H)} term in the above equation can then be replaced with the equivalent \textit{instantaneous} bolometric luminosity ($L_{bol}$),\\
\begin{equation}
 Q(H) = \frac{L_{bol}}{h\nu} 
\end{equation}
This bolometric luminosity can then be replaced with a crude assumption, i.e. $L_{bol} = A\times \mathrm{L_{5100}}$, where \textit{A} is the bolometric correction \citep[see][ for a recent review]{netzer2019}. This gives us the relation,\\
\begin{equation}
    \log U = \log \left ( \frac{A\mathrm{L_{5100}}}{h\nu_{5100}\times 4\pi R_{BLR}^2 n_H c}\right )
    \label{u-n}
\end{equation}
From the \textit{Clean2} sample of \cite{bentz13}, I have,\\
\begin{equation}
    \log \left ( \frac{R_{BLR}}{1\;light-day}\right )  = \kappa + \alpha \log\left ( \frac{\mathrm{L_{\lambda}}}{10^{44}}\right )
    \label{r-l}
\end{equation}

Substituting Equation \ref{r-l} in Equation \ref{u-n}, I have
\begin{equation}
\begin{split}
    \log U = & \log \left [ \frac{A}{4\pi h\nu_{\lambda} (1\;light-day)^2 c}\right] -2(\kappa - 44\alpha)\\
   & + \left [(1-2\alpha)\log \mathrm{L_{\lambda}}\right ] - \log n_H
\end{split}
\end{equation}

Here, the values for $\kappa$ and $\alpha$ used are 1.555 and 0.5 (instead of the quoted value of 0.542), respectively. Also, I assume an average value for A = 9 \citep{richards06, elvis94}. I get a simplified relation between \textit{U} and \textit{$\rm{n_H}$},\\
\begin{equation}
    \log U = B - \log n_H
\end{equation}
Here, the value of B = 10.85$\bar{1}$, where B is\\
\begin{equation}
    B = \log \left (\frac{A}{4\pi h\nu_{5100}c}\right ) - 2\kappa + \log 10^{44} -2\log (1\;light-day)
\end{equation}
\par 


\subsubsection{In terms of $\lambda_{Edd}$ and $M_{BH}$}
\label{exten}
Interpreting this in terms of \textit{Eddington ratio}, $\lambda_{Edd}$\\
\begin{equation}
    \log U = C - \log n_H + (1-2\alpha)\log \left [ \lambda_{Edd} L_{Edd}\right]
    \label{ledd}
\end{equation}
where,
\begin{equation}
    C = B - (1-2\alpha)\log A
\end{equation}
Equation \ref{ledd} can then be re-written in terms of black-hole mass, \textit{$M_{BH}$}
\begin{equation}
    \boxed{\log U = D - \log n_H + (1-2\alpha)\log \left [ \lambda_{Edd} M_{BH}\right]}\\ 
\end{equation}
where,
\begin{equation}
    D = C + (1-2\alpha)\log \left ( \frac{4\pi GM_{\odot}m_{p}c}{\sigma_{T}}\right )
\end{equation}
where, \textit{$M_{BH}$} is measured in units of solar mass ($M_{\odot}$); \textit{G} is the Gravitational constant; $m_p$ is the mass of a proton (in cgs); $\sigma_{T}$ is the Thompson's cross-section (in cgs).\\

In Panda et al. (in prep.), I combine this knowledge with two other key entities -- (a) the departure coefficient $\Delta R_{H\beta}$ \citep{martinez-aldama2019}; and (b) the BLR fundamental plane relation \citep{dupu2016L}. Eventually, I have an analytical expression that combines the BLR picture from the physical point of view, i.e., using the ionization parameter (\textit{U}) and cloud density (\textit{n$_{\rm{H}}$}), and, to the observational perspective, \rfe{} and \LLEdd{}. This additionally will include the dependence on the shape of the \hb{} line profile and the effect of the viewing angle. 

\subsection{CLOUDY simulations}
\label{sub:cloudy}
Taking inspiration from the Locally Optimally Emitting Clouds (LOC) model \citep{baldwin1995}, I perform a suite of models by varying the cloud particle density, \textit{n$_{\rm{H}}$}, and the ionization parameter, \textit{U}. The model assumes a distribution of cloud densities at various radii from the central illuminating source to mimic the gas distribution around the close vicinity of the active nuclei. Although, I extract directly the emission line information from each of the single cloud models and do not integrate the line emission from the clouds. The remaining parameters are the cloud column density (\textit{N$_{\rm{H}}$}) for which I incorporate a value of $10^{24}\;\rm{cm^{-2}}$ motivated by our past studies \citep{panda_frontiers,panda18b,panda19}. Indeed, one expects a broad range of column densities to be present in the BLR, yet, this value of the \textit{N$_{\rm{H}}$} quite consistently reproduces the observed line emission, especially in the case of the optical and UV \feii{} as is shown in \citet{bv08}. I use solar abundance which are estimated using the \textit{GASS10} module \citep{gass10}. 

I utilize the spectral energy distribution for the nearby (z= 0.0611) Narrow Line Seyfert 1 (NLS1), \textit{I Zwicky 1} (hereafter I Zw 1) (Also known by PG 0050+124 or Mrk 1502. The I Zw 1 ionizing continuum shape is obtained from \href{http://vizier.u-strasbg.fr/vizier/sed/}{http://vizier.u-strasbg.fr/vizier/sed/}). I Zw 1 is the prototypical optical NLS1, with strong \feii{} emission and unusually narrow permitted lines, e.g.  H$\beta$ FWHM 1240 \kms{} \citep{oster85} and it is also a luminous radio-quiet PG QSO (M$\rm{_B}$=-23.5, \citealt{pg83}). The bolometric luminosity of I Zw 1 is L$_{\rm{bol}} \sim 3\times10^{45}$ erg s$^{-1}$ \citep{porquet2004}, which for a black hole mass of 2.8$^{+0.6}_{-0.7}\times 10^7\;\rm{M_{\odot}}$ \citep{vester06}, implies that I Zw 1 accretes at close to the Eddington limit. The parameter \rfe\ is extracted from these simulations.

\section{Results and Discussions}
\label{sub:results}

It has been shown in previous studies \citep{adhikari18} that there is a substantial change in the gas pressure profile and correspondingly the density profiles when models with constant density and constant pressure are compared side-by-side. The profiles are almost in-line for lower densities i.e., n$_{\rm{H}}$ $\lesssim$ 10$^{9}$ cm$^{-3}$. But, the profiles start to diverge for n$_{\rm{H}}$ $\gtrsim$ 10$^{9}$ cm$^{-3}$ and then substantially becomes wider. Thus, for dense clouds ($\rm{n_H} \gtrsim 10^{10} \rm{cm}^{-3}$), constant pressure and constant density models give very similar \rfe{} estimates.

\begin{figure}
    \centering
    \includegraphics[width=\columnwidth]{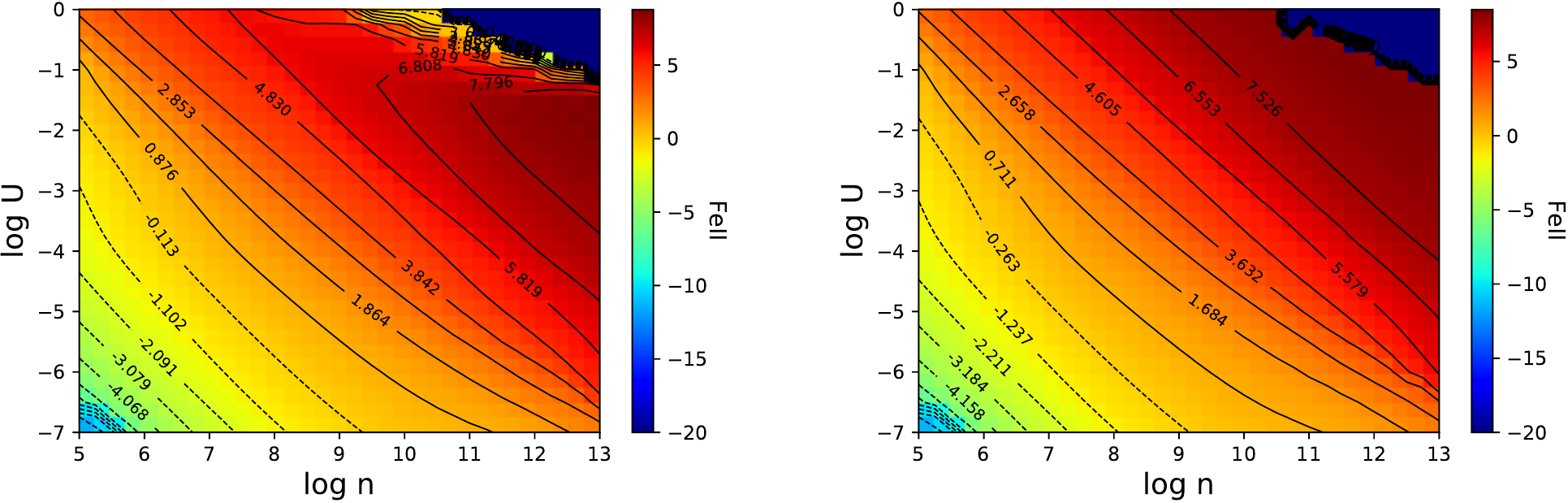}
    \caption{2D histograms showing log U - log $\rm{n_H}$ for \feii{} for (a) constant density; (b) constant pressure. The model uses solar abundance (\zsun{}) and 0 \kms{} microturbulence.}
    \label{fig:feii}
\end{figure}

\begin{figure}
    \centering
    \includegraphics[width=\columnwidth]{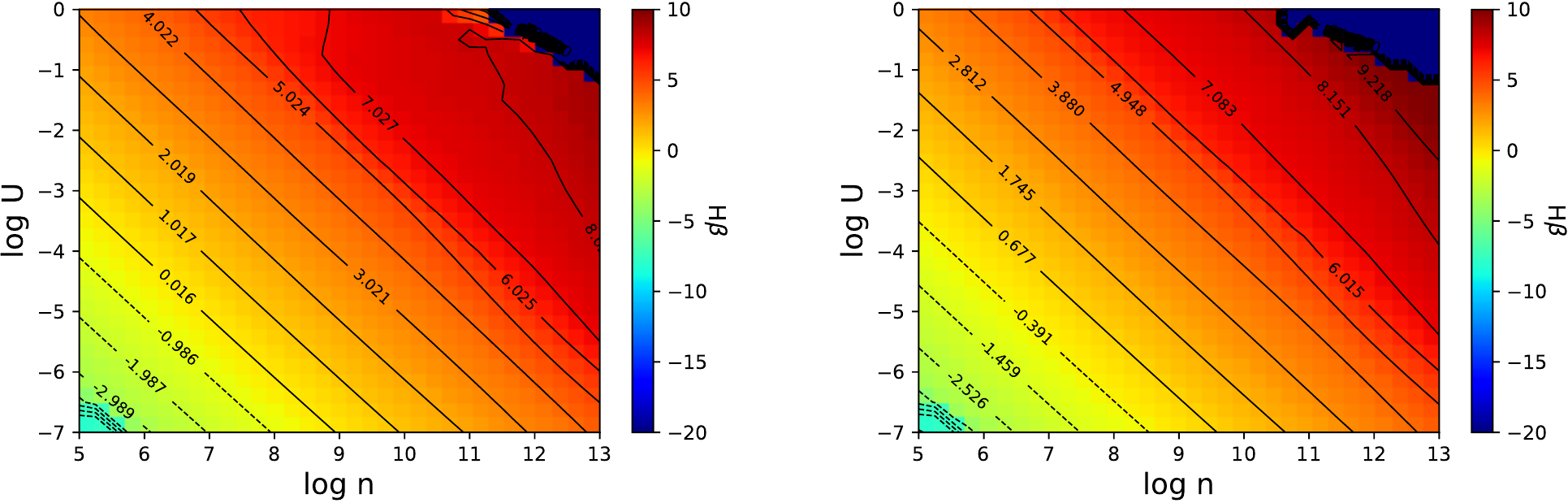}
    \caption{2D histograms showing log U - log $\rm{n_H}$ for \hb{} for (a) constant density; (b) constant pressure. The model uses solar abundance (\zsun{}) and 0 \kms{} microturbulence.}
    \label{fig:hb}
\end{figure}

\begin{figure}
    \centering
    \includegraphics[width=\columnwidth]{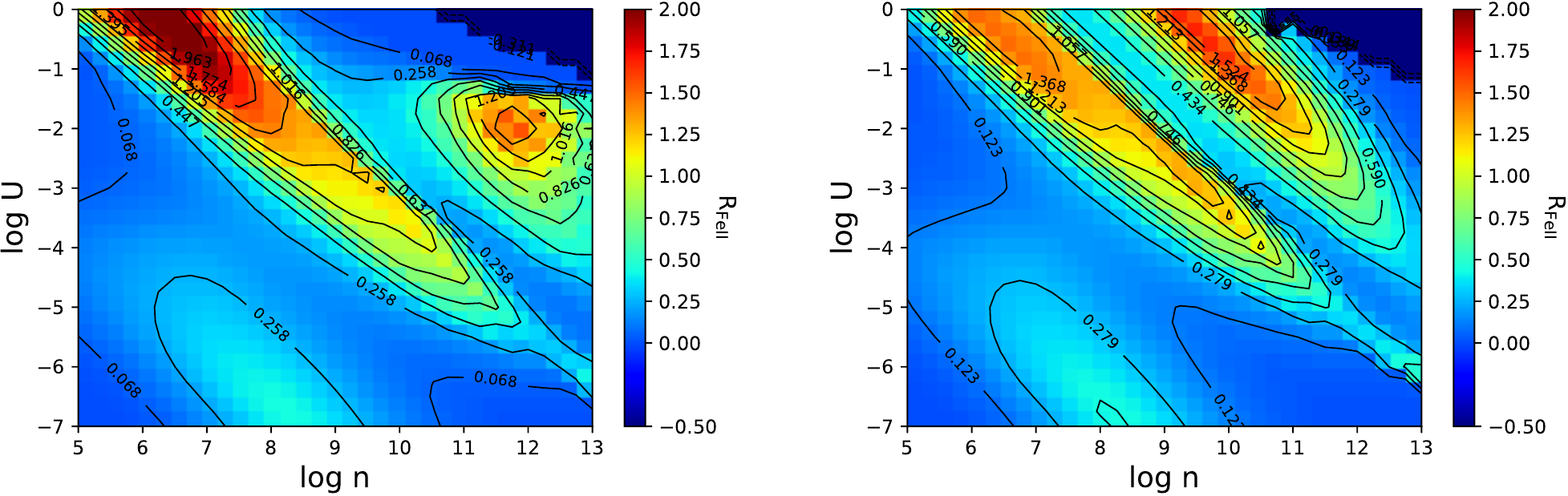}
    \caption{2D histograms showing log U - log $\rm{n_H}$ for \rfe{} for (a) constant density; (b) constant pressure. The model uses solar abundance (\zsun{}) and 0 \kms{} microturbulence.}
    \label{fig:rfe}
\end{figure}

In the context of the current work, I show this difference in a more extended log U - log $\rm{n_H}$ parameter space and comparing the two models side-by-side (Figures \ref{fig:feii}, \ref{fig:hb} and \ref{fig:rfe}). The results shown in both the panels in Figures \ref{fig:feii}, \ref{fig:hb} and \ref{fig:rfe} are without the effect of dust. This effect will be presented and analysed in an upcoming paper. The density maps clearly show multiple peaks in the \rfe{} values as a function of changing log U and log $\rm{n_H}$, especially the three peaks -- (Region A) short region about the (-6.5,8); (Region B) the elongated region in the middle spanning from (-6,13) to (0,5); and (Region C) third patch that extends from (-4.5,13) to (-1,9). Taking hint from various previous works, the effective BLR densities have been found to span between $\sim10^{9}$--$10^{12}$ (in $\;\rm{cm}^{-3}$) or even higher in some cases \citep{dragana2012, 2014MNRAS.438..604B, marz15, 2019arXiv190800742C}. Additionally, in the ionisation parameter context, very high values imply that the BLR clouds can be maximally ionized and this will lead to suppression of various line emission. This will not only quench the observed emission from HILs but also for the LILs such as the H$\beta$ and the \feii{} emission observed in the optical band. Thus, the region A and almost whole of the region B gets omitted based on these reasoning. In Panda et al. (in prep.), I show that with the inclusion of dust (applying realistic dust temperature prescription from \citealt{Nenkova2008}) we are finally left with the region C which best describes the BLR in AGNs.

\section{Conclusions and Future}
\label{sub:future}

I derive an analytical expression that will seamlessly tie together the fundamental BLR properties coming from theory and observations. This relation combines the \un{} relation from the photoionization theory with the observationally constrained relations, such as the \RL{} (and others as described in Sec. \ref{exten}), in terms of the Eigenvector 1 parameter, \rfe{} \citep[for more details see][]{panda18b, panda19, panda19c, panda19b}. I have shown preliminary results from a suite of models using CLOUDY photoionization code. These results will be further expanded and tested for -- (i) the effect of dust inclusion at larger depths within the BLR clouds; (ii) the effect of changing the ionizing continua; and (iii) changing chemical composition and dynamics within the cloud.

\acknowledgements{The project was partially supported by National Science Centre, Poland, grant No. 2017/26/A/ST9/00756 (Maestro 9). I would like to thank Bo\.zena Czerny, Mary Loli Mart\'inez Aldama, Paola Marziani, Deepika Ananda Bollimpalli and Abbas Askar for fruitful discussions leading to this project.}

\bibliographystyle{ptapap}
\bibliography{panda1}

\end{document}